\documentclass[epjCONF,columns]{svjour} 
\usepackage{graphics}
\usepackage[varg]{txfonts} 
\usepackage[latin1]{inputenc}
\session-title{2011 Hadron Collider Physics symposium (HCP-2011),
     Paris, France, November 14-18 2011}
\begin{document}
\title{Heavy flavor physics with the CMS experiment}
\author{Vincenzo Chiochia\thanks{\email{vincenzo.chiochia@cern.ch}} \\{\it on behalf of the CMS Collaboration} }
\institute{Physik-Institut, Universit\"at Z\"urich, Winterthurerstr. 190, CH-8057 Zurich, Switzerland}
\abstract{Thanks to the excellent tracking and muon identification performance, combined with a flexible trigger system, the CMS experiment at the Large Hadron Collider is conducting a rich and competitive program of measurements in the field of heavy flavor physics. We review the status of b-quark production cross section measurements in inclusive and exclusive final states, the measurement of B hadron angular correlations, the search for rare $B_{s}^0$ and $B^0$ decays to dimuons, and the observation of the $X(3872)$ resonance.
} 
\maketitle
%
%
\section{Introduction\label{sec:intro}}

The Compact Muon Solenoid (CMS) experiment at the Large Hadron Collider (LHC) has a rich and competitive hea\-vy flavor program including measurements of b-quark production, studies of B hadron decays, searches for B baryons as well as measurements of quarkonium and exotic states production. Indirect searches for new physics, such as the  rare $B_{s,d}^0 \rightarrow \mu^+ \mu^-$ decays, studies of $b \rightarrow s \,\mu^+ \mu^-$ transitions, and measurements of CP-violating phases in the $B_s$ sector, provide important constraints to the Standard Model (SM) and are complementary to direct searches. 

In 2010 the CMS experiment has collected an integrated luminosity of 40~pb$^{-1}$ at $\sqrt{s}=7$~TeV with peak instantaneous luminosities of $2 \times 10^{32}$~cm$^{-2}$s$^{-1}$. The 2011 data taking period was characterized by a steep increase of the LHC instantaneous luminosity. A peak instantaneous luminosity of $3.5 \times 10^{33}$~cm$^{-2}$s$^{-1}$ was reached towards the end of the run and the total integrated luminosity recorded was about 5~fb$^{-1}$.
The main ingredients for the CMS heavy flavor program are di-muon triggers combined with precise tracking and vertex reconstruction capabilities. The flexibility of the CMS trigger system has made it possible to adapt the online event selection algorithms to the increasing luminosity in a prompt and intelligent manner, by making use of invariant mass, decay length, distance of closest approach, transverse momentum, and rapidity in the trigger selection criteria. Due to bandwidth limitation the scope of the CMS heavy flavor program was limited to final states with di- or multi-muons and b-tagged jets.

This article is structured as follows: in Section~\ref{sec:bprod} we summarize the status of b-quark production measurements; in Section~\ref{sec:bcorr} we present the measurements of B hadron angular correlations; in Section~\ref{sec:raredecays} we review the recent searches for rare $B_s^0$ and $B^0$ decays to dimuons; in Section~\ref{sec:X3872} we present the observation of the $X(3872)$ resonance and the conclusions are given in Section~\ref{sec:concl}.

%
\section{Measurements of b-quark production\label{sec:bprod}}

Understanding b-quark production at the LHC experiments is important for various reasons. Firstly, cross section measurements are required to test QCD predictions, which have been computed at next-to-leading order (NLO) precision but are still characterized by large scale dependence. Secondly, final states with b-jets represent an important source of background for many of the most interesting physics searches, such as the Higgs boson and extensions of the Standard Model.

Production of b-quarks at $\sqrt{s}=7$~TeV was measured by the CMS experiment in both inclusive and exclusive channels using pp collision data collected in 2010. Inclusive measurements at low instantaneous LHC luminosities were initially done using final states with a single muon~\cite{Khachatryan:2011hf}, thanks to the relatively low muon trigger thresholds, and with b-tagged jets~\cite{CMS-PAS-BPH-10-009}. These results are summarized in Ref.~\cite{Chiochia:2010aw}. More recently, further cross section measurements were carried out using di-muon final states~\cite{CMS-PAS-BPH-10-015}, events with muons plus b-tagged jets~\cite{CMS-PAS-BPH-10-008}, and from non-prompt $J/\psi \rightarrow \mu^+ \mu^-$ decays~\cite{Khachatryan:2010yr,Chatrchyan:2011kc}. 
\begin{table*}[htdp]
\begin{center}
\begin{tabular}{lccccc} \hline
Process & $p_T$ range  & (Pseudo)-rapidity  & Cross section & NLO QCD  & Ref. \\ 
 & [GeV/c] & range & [$\mu$b] & [$\mu$b] & \\ \hline 
\hline 
$pp \rightarrow b X \rightarrow \mu X$ & $6-\infty$ & $|\eta|<2.1$ & $1.32\pm 0.01\pm 0.30\pm 0.15$ & $0.95^{+0.41}_{-0.21}$ & \cite{Khachatryan:2011hf}\\
$pp \rightarrow b\bar{b} X \rightarrow \mu\mu X$ & $4-\infty$ & $|\eta|<2.1$ & $(26.18\pm 0.14\pm 2.82\pm 1.05)\times 10^{-3}$ & $(19.95^{+4.68}_{-4.33})\times 10^{-3}$ & \cite{CMS-PAS-BPH-10-015}\\
$pp \rightarrow {\rm b~jet}~X$ & $30-\infty$ & $|y|<2.4$ & $2.14\pm 0.01\pm 0.41\pm 0.09$ & $1.83^{+0.64}_{-0.42}$ & \cite{CMS-PAS-BPH-10-008}\\
\hline
$pp\rightarrow B^+ X$ & $5-\infty$ & $|y|<2.4$ & $28.3\pm 2.4\pm 2.0\pm 1.1$ & $25.5^{+8.8}_{-5.4}$ & \cite{Khachatryan:2011mk}\\ 
$pp\rightarrow B^0 X$ & $5-\infty$ & $|y|<2.2$ & $33.2\pm 2.5\pm 3.1\pm 1.3$ & $25.2^{+9.6}_{-6.2}$ & \cite{Chatrchyan:2011pw} \\
$pp\rightarrow B_s X \rightarrow J/\psi\phi X$ & $8-50$     & $|y|<2.4$ & $(6.9\pm 0.6\pm 0.5\pm 0.3)\times 10^{-3}$ & ($4.9^{+1.9}_{-1.7}) \times 10^{-3}$ & \cite{Chatrchyan:2011vh}\\
\hline
\end{tabular}
\end{center}
\caption{Summary of measured b-quark cross sections in inclusive (top) and exclusive (bottom) channels and comparison with NLO QCD expectations. The experimental uncertainties indicate (from left to right) the statistical, systematic and luminosity. Theoretical uncertainties are obtained from scale variations.\label{tab:bxsec}}
\end{table*}%

\begin{figure*}[htbp]
\begin{center}
\resizebox{0.67\columnwidth}{!}{\includegraphics{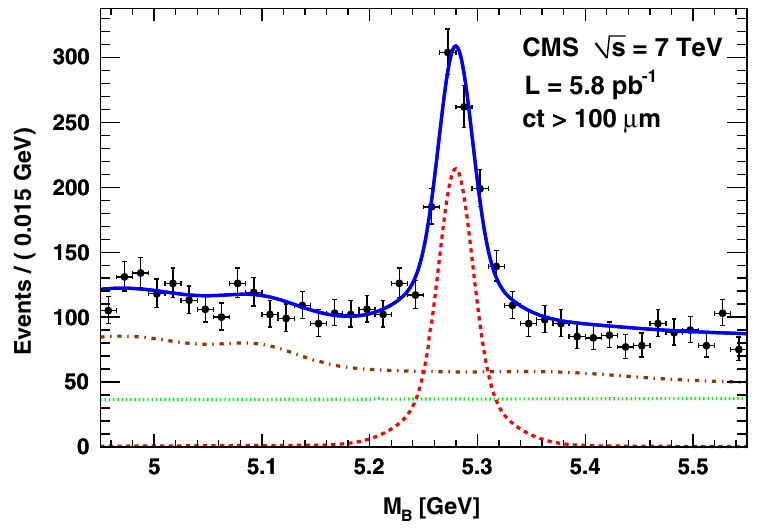} }
\resizebox{0.70\columnwidth}{!}{\includegraphics{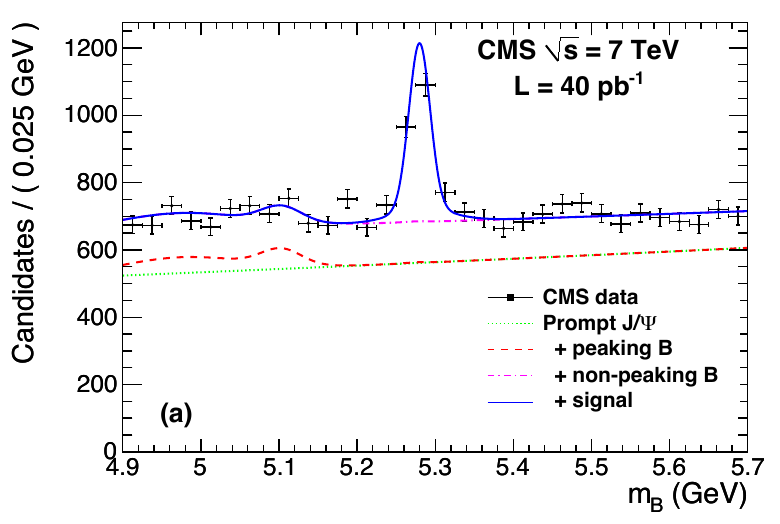} }
\resizebox{0.66\columnwidth}{!}{\includegraphics{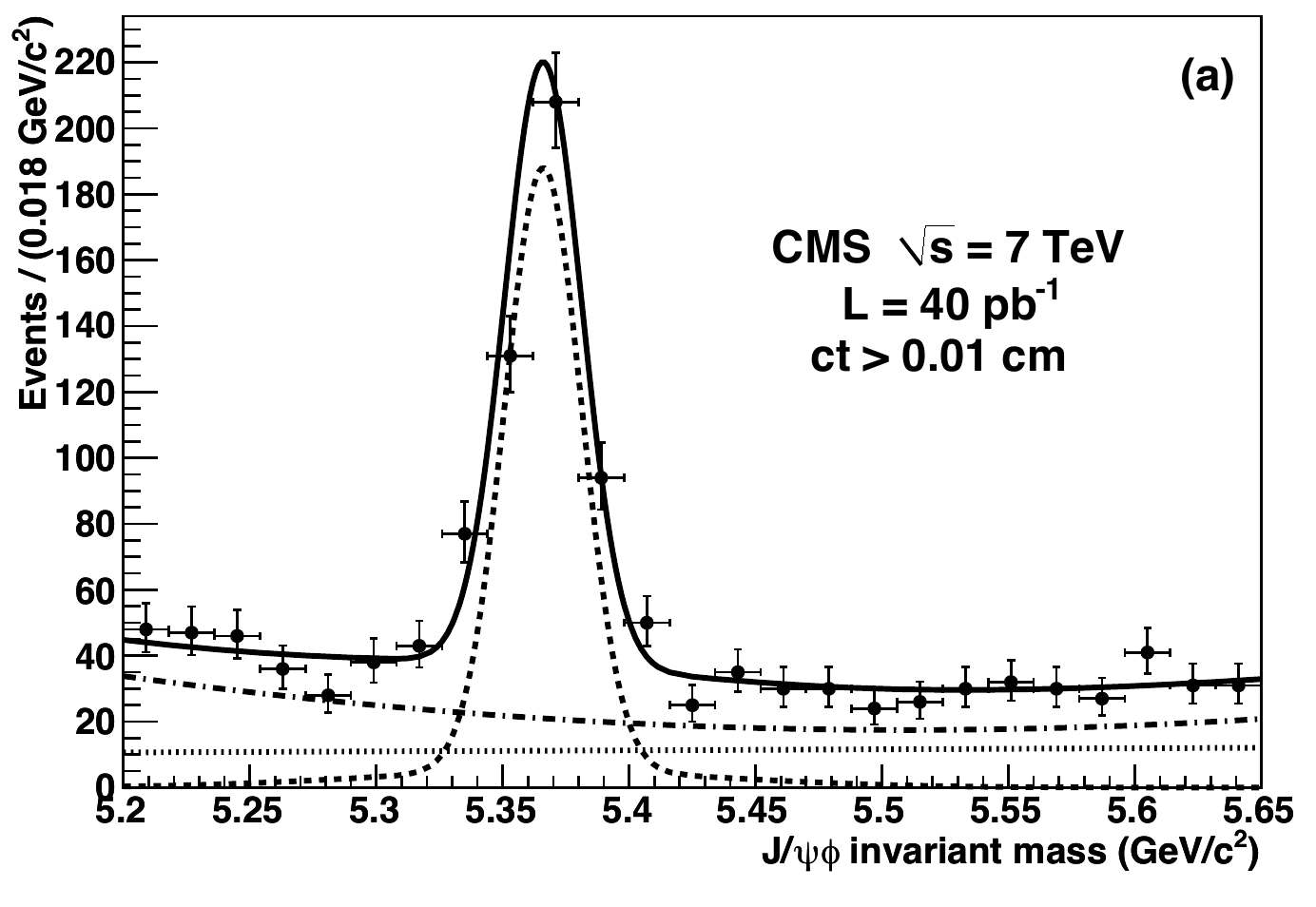} }
\caption{Invariant mass distributions of the decays $B^+ \rightarrow J/\psi K^+$ (left), $B^0 \rightarrow J/\psi K^0_s$ (center) and $B_s \rightarrow J/\psi \phi$ (right). The CMS data points are represented by the full dots. The solid lines are fits to the data.\label{fig:Bmassplots}}
\end{center}
\end{figure*}

Differential cross section measurements versus transverse momentum and rapidity were also performed using fully-reconstructed decays of B mesons, such as $B^+ \rightarrow J/\psi K^+$, $B^0 \rightarrow J/\psi K_s^0$ and $B^0_s \rightarrow J/\psi \phi$~\cite{Khachatryan:2011mk,Chatrchyan:2011pw,Chatrchyan:2011vh}. In particular, the measurement of the $B_s$ production cross section in the $J/\psi \phi$ decay is an important first step towards establishing the CP violating phase in the $B_s$ system. The analyzed integrated luminosity ranged between 6 and 40~pb$^{-1}$. $J/\psi$ decays were reconstructed in the dimuon final state and then combined with one or two hadron tracks from the same secondary vertex, or with a displaced tertiary vertex from a $K_s^0 \rightarrow \pi^+ \pi^-$ decay. The fractions of signal and background events were extracted from a combined fit to the invariant mass and lifetime distributions, as shown in Figure~\ref{fig:Bmassplots}. 

\begin{figure}[htbp]
\begin{center}
\resizebox{0.95\columnwidth}{!}{\includegraphics{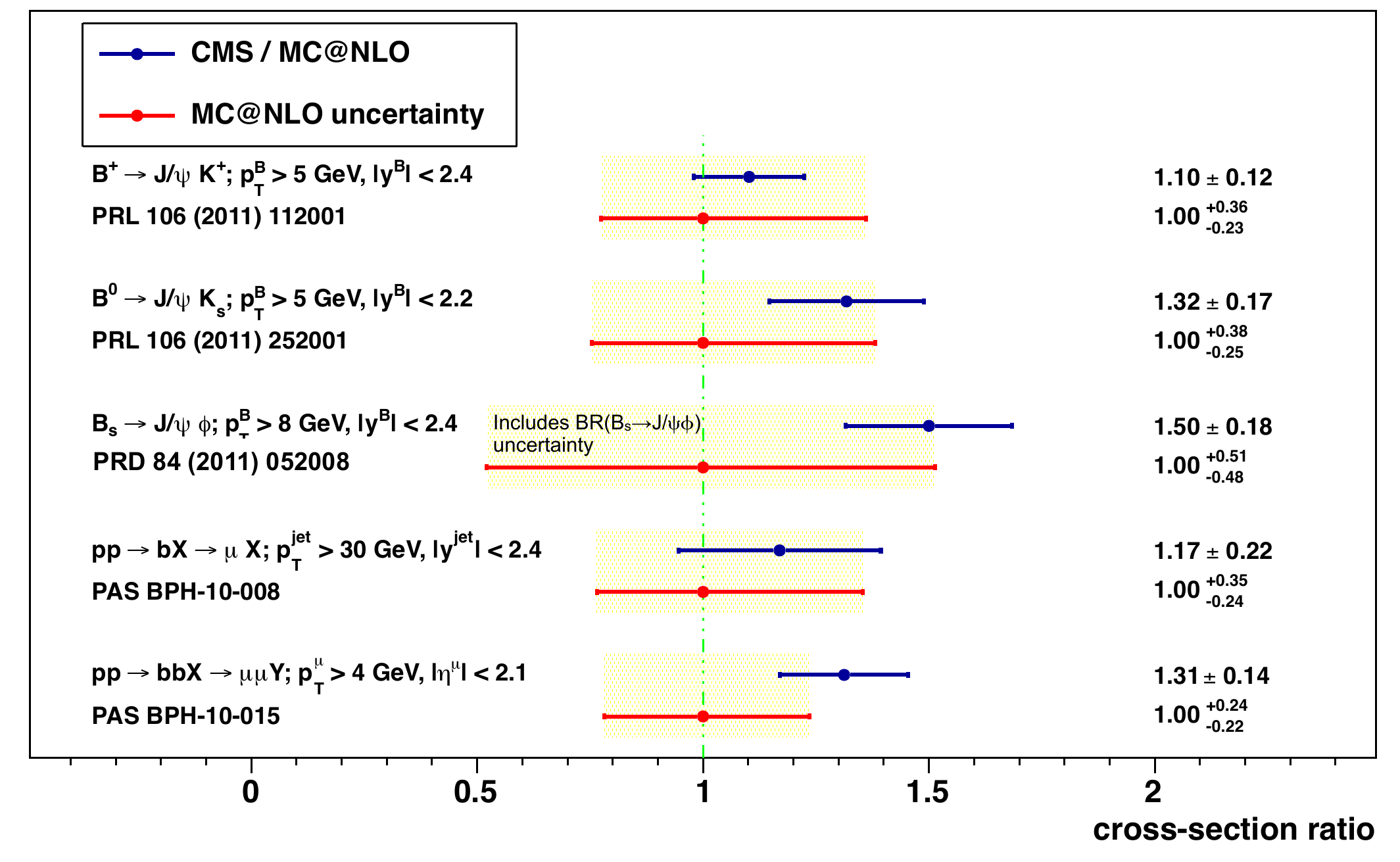} }
\caption{Ratio of measured and predicted b-quark cross sections.\label{fig:dataNLOratio}}
\end{center}
\end{figure}

The  cross sections, integrated over the measured $p_T$ and $y$ ranges, are summarized in Table~\ref{tab:bxsec}, along with the NLO QCD predictions obtained with the {\tt MC@NLO} Monte Carlo (MC) generator. The ratios of measured over predicted NLO cross sections are shown in Fig.~\ref{fig:dataNLOratio}. The measurements, probing b-quark production in different kinematic regions, are in agreement with the NLO QCD expectations although NLO predictions are often below the measurements. Theoretical uncertainties are always larger than the experimental ones and are dominated by scale uncertainties.


\section{Measurements of B hadron angular correlations\label{sec:bcorr}}

In the lowest order perturbative QCD momentum conservation requires the $b$ and $\bar{b}$ quarks to be emitted in a back-to-back topology. However, higher order subprocesses with additional emitted partons, such as gluons, give rise to different topologies of the final state b quarks. Con\-se\-quen\-tly, measurements of $b\bar{b}$ angular and momentum correlations provide information about the underlying production subprocesses and allow for a sensitive test of leading-order  and  NLO QCD cross sections and their evolution with event energy scales.

\begin{figure}[htbp]
\begin{center}
\resizebox{\columnwidth}{!}{\includegraphics{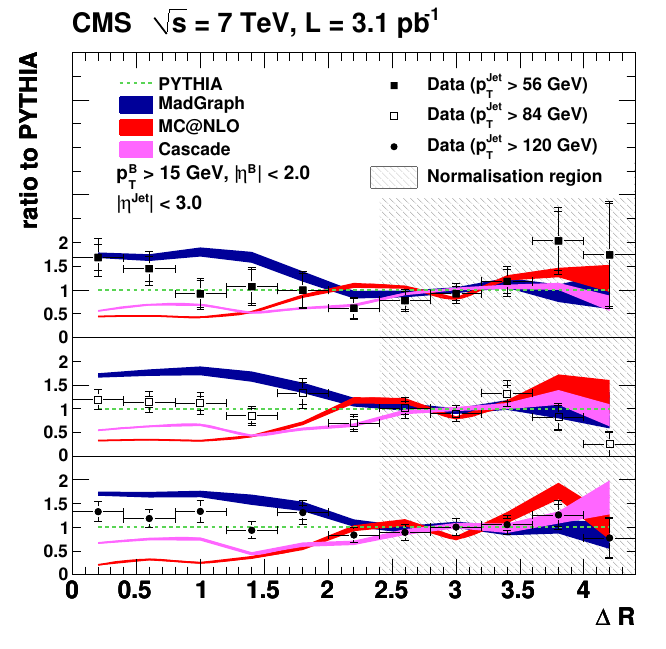} }
\caption{Ratio of the differential B hadron pair production cross sections, as a function of $\Delta R$ for data and various Monte Carlo generators with respect to the {\tt PYTHIA} predictions, for  three leading jet $p_T$ regions. The simulation is normalised to the region $\Delta R > 2.4$, as indicated by the shaded normalisation region. \label{fig:BBcorrelations}}
\end{center}
\end{figure}

The angular correlations of B hadron pairs were recently measured with the CMS experiment using a novel technique based on se\-con\-da\-ry vertices and accessing for the first time the region of collinear b-quark pair production~\cite{Khachatryan:2011wq}. 
An inclusive secondary vertex finding (IVF) technique, completely independent of jet reconstruction, is applied for this purpose. This technique reconstructs se\-con\-da\-ry vertices (SV) by clustering tracks around the so-called {\it seeding tracks} characterized by high three-dimensional impact parameter significance. The four-momentum of the reconstructed B hadron candidate is then identified with the SV four-momentum. The angular separation between B candidate pairs is measured for three different regions of the leading jet $p_T$ in the event. The ratios of measured and simulated cross sections over the {\tt PYTHIA} predictions are shown in Figure~\ref{fig:BBcorrelations}. The data lie between the {\tt MADGRAPH} and the {\tt PYTHIA} MC predictions. Neither the {\tt MC@NLO} nor the {\tt CASCADE} calculations describe the shape of the $\Delta R$ distribution well. In particular the collinear region at small values of $\Delta R$, where the contributions of gluon splitting processes are expected to be large, is not adequately described by any of the predictions. In addition, the fraction of cross section in this collinear region is found to increase with the leading jet $p_T$.

%
%
\section{Searches for rare B hadron decays\label{sec:raredecays}}

The rare decays $B_{s,d}^0 \rightarrow \mu^+ \mu^-$ are extremely interesting for new physics searches. The decays are flavour-changing neutral current processes which are forbidden in the SM at a tree level, occurring only via higher order diagrams. 
The SM-predicted branching fractions, Br$(B_s \rightarrow \mu^+ \mu^-)=(3.2\pm 0.2)\times 10^{-9}$ and Br$(B^0 \rightarrow \mu^+ \mu^-)=(1.0\pm 0.1)\times 10^{-10}$~\cite{Buras:2010wr}, are significantly enhanced in several extensions of the SM, although in some cases the decay rates are lowered~\cite{Ellis:2007kb}.

At the LHC the search for these decays was conducted both by the CMS~\cite{Chatrchyan:2011kr} and LHCb collaborations~\cite{LHCb:2011ac}. In the recent LHCb result, based on 0.37~fb$^{-1}$, the number of events in the signal window was compatible with the sum of background and SM signal. The upper limit on the branching ratio at 95\% C.L. was $1.6 \times 10^{-8}$ and $3.6 \times 10^{-9}$, for the $B_s$ and $B^0$ decays, respectively.

The first CMS search is based on 1.14~fb$^{-1}$ collected in 2011. An event-counting experiment is performed in dimuon mass regions around the $B_s^0$ and $B^0$ masses. Two additional event samples are selected. A sample of about 17\,000 $B^+ \rightarrow J/\psi K^+$ events is used as a normalization sample to minimize uncertainties related to the b-production cross section and to the integrated luminosity. A sample of $B_s^0 \rightarrow J/\psi \phi$ is used to validate the MC simulation and to evaluate potential effects resulting from differences in fragmentation between $B^+$ and $B^0_s$. More details of the CMS event selection can be found in Ref.~\cite{Martini} of these conference proceedings.

Events in the signal window can result from real signal decays, combinatorial background, and backgrounds which tend to peak in the dimuon mass. The latter are from decays of the type $B \rightarrow h h'$, where $h$ and $h'$ are charged hadrons misidentified as muons. The expected number of combinatorial background events is evaluated by interpolating to the dimuon mass signal window the number of events observed in the sideband regions. The expected number of peaking background events is evaluated from MC and using muon misidentification rates measured from data.

Three and one events are observed in the $B_s$ and $B^0$ mass windows, respectively. The number of observed events is consistent with the SM expectation for signal plus background and upper limits on the branching ratios are determined with the $\rm{CL}_s$ approach. The obtained limits at 95\% C.L. are $1.9 \times 10^{-8}$ and $4.6 \times 10^{-9}$ for the $B_s$ and $B_0$ decays, respectively. The combination of CMS and LHCb upper limits for the $B^0_s$ decay gives $1.1 \times 10^{-8}$ at 95\% C.L.~\cite{Bscombination}. This result provides strong constraints on supersymmetric models at large values of $\tan \beta$. The analysis of the full 2011 dataset was ongoing at the time of writing. 
\section{Observation of the X(3872) resonance\label{sec:X3872}}

The $X(3872)$ resonance was discovered in 2003 by the Belle collaboration and later confirmed by BaBar, Belle and D0. The quantum numbers $J^{PC}$ of the $X(3872)$ state are not yet fully established. A study of the angular decay distributions performed by the CDF Collaboration gave non-negligible probabilities only for $1^{++}$ and $2^{-+}$~\cite{Abulencia:2006ma}. 
Despite a large experimental effort, the nature of this new state is still uncertain and several models have been proposed to describe it. The $X(3872)$ could be a conventional charmonium state, a loosely bound $D^{*0}\bar{D}^0$ molecule or a tetraquark state. Measurements of the production cross section at the LHC may help in disentangling the nature of this resonance.

\begin{figure}[htbp]
\begin{center}
\resizebox{1.0\columnwidth}{!}{\includegraphics{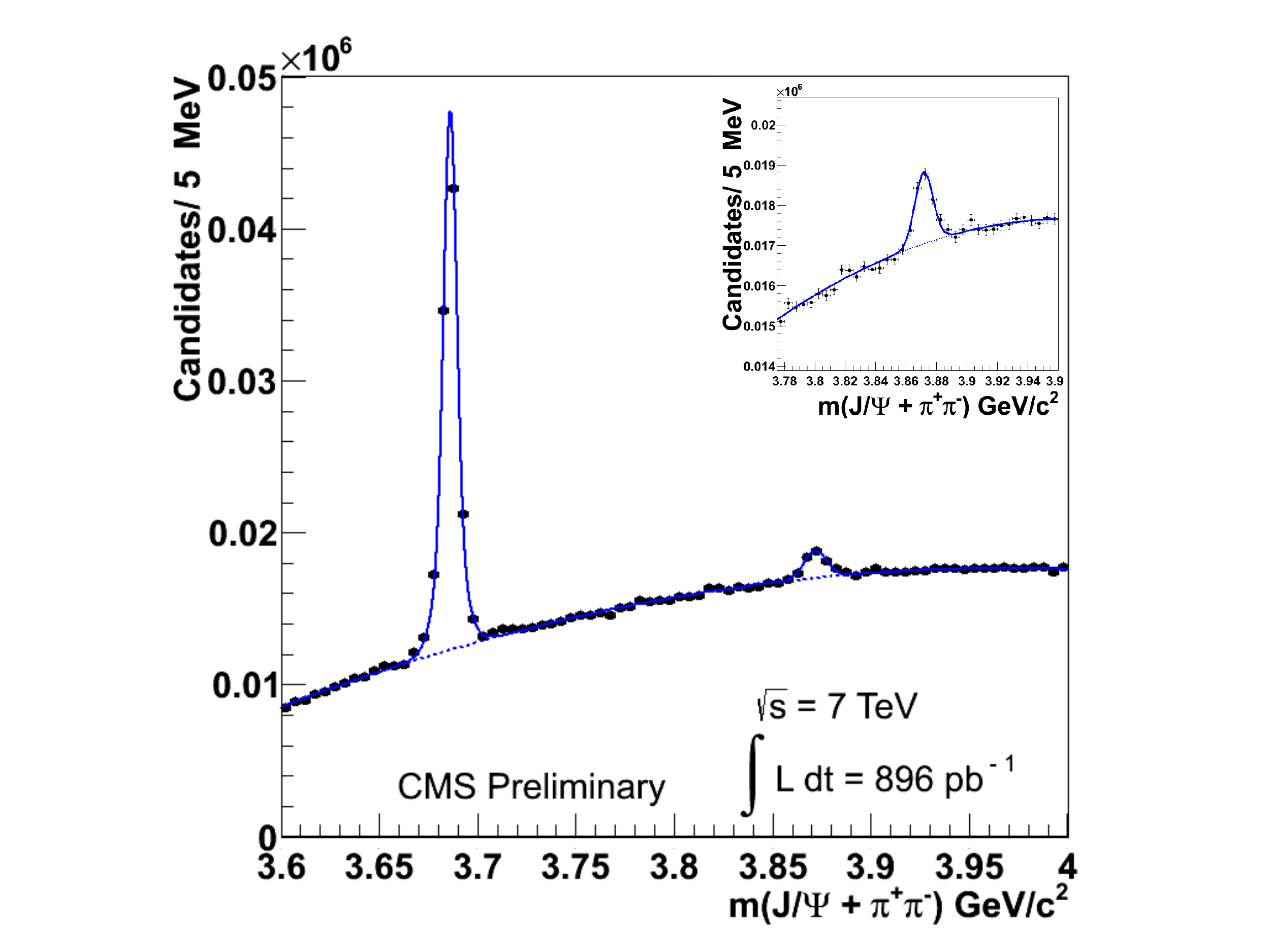} }
\caption{$J/\psi \pi^+ \pi^-$ invariant mass spectrum in the region $p_T(X) > 8$~GeV and $|y(X)| < 2.2$. The insert shows the mass region around the $X(3872)$.\label{fig:X3872}}
\end{center}
\end{figure}

First observations of this state at the LHC were reported by both CMS~\cite{CMS-PAS-BPH-10-018} and LHCb collaborations~\cite{Aaij:2011sn} in the decay to $J/\psi \pi^+ \pi^-$, using data collected in 2010. The CMS collaboration performed a measurement of the ratio of the production cross sections for the $X(3872)$ and the $\psi (2S)$ resonances in the kinematic region defined by $p_T(X) > 8$~GeV and $|y(X)| < 2.2$. The differences in acceptance and efficiency were accounted for by correction factors determined from Monte Carlo simulations. The $X(3872) $ signal yield of $548\pm 104$ events was extracted from an unbinned log likelihood fit to the invariant mass spectrum of the $J/\psi \pi^+ \pi^-$ system. The measured cross section ratio is $R = 0.087\pm 0.017\pm 0.009$, where the first error refers to the statistical uncertainty of the data and the second refers to the systematic uncertainty. The search for the $X(3872)$ resonance was updated using a larger data set collected in 2011 and about 5\,300 candidates were observed in 896~pb$^{-1}$~\cite{CMS-DPS-2011-009}, as shown in Figure~\ref{fig:X3872}.

\section{Conclusions\label{sec:concl}}

The CMS experiment has a rich program of studies in the field of heavy flavor physics. The b-quark production cross section was measured in several inclusive and exclusive channels and compared to QCD predictions. Angular correlations of B hadron pairs were measured using a novel technique based on secondary vertices, and accessing the collinear region for the first time. Searches for the rare $B_s$ and $B^0$ decays to dimuons have been conducted, setting stringent constraints on extensions to the Standard Model. In addition, the resonance $X(3872)$ was observed and a preliminary measurement of the production cross section ratio over the $\psi(2S)$  was performed. 

%

\end{document}